\documentclass[prb,preprint,amsmath,amssymb]{revtex4}

\usepackage{graphicx}
\usepackage{dcolumn}
\usepackage{bm}

\begin{document}

\title{Development of a Classical Force Field for the
Oxidised Si Surface: Application to Hydrophilic Wafer Bonding}

\author{Daniel J. Cole}
\email{djc56@cam.ac.uk}
\affiliation{Theory of Condensed Matter Group, Cavendish Laboratory, \\
           University of Cambridge, J J Thomson Avenue, Cambridge, CB3 0HE, UK}
\author{G\'{a}bor Cs\'{a}nyi}
\affiliation{Centre for Micromechanics, Department of Engineering, \\
           University of Cambridge, Trumpington Street, Cambridge, CB2 1PZ, UK}
\author{S. Mark Spearing}
\affiliation{Engineering Materials Group, School of Engineering Sciences,\\
            University of Southampton, Southampton, SO17 1BJ, UK}
\author{Mike C. Payne}
\affiliation{Theory of Condensed Matter Group, Cavendish Laboratory, \\
           University of Cambridge, J J Thomson Avenue, Cambridge, CB3 0HE, UK}
\author{L. Colombi Ciacchi}
\affiliation{Fraunhofer Institut f\"ur Werkstoffmechanik \\
           W\"ohlerstrasse 11, 79108 Freiburg, Germany}
\affiliation{Institut f\"ur Zuverl\"assigkeit von Bauteilen und Systemen, \\
           University of Karlsruhe, Kaiserstr. 12, 76131 Karlsruhe, Germany.}

\date{\today}

\begin{abstract}
We have developed a classical two- and three-body interaction potential to simulate the 
hydroxylated, natively oxidised Si surface in contact with water solutions, based on 
the combination and extension of the Stillinger-Weber potential and of a potential 
originally developed to simulate SiO$_{2}$ polymorphs.
The potential parameters are chosen to reproduce the structure, charge distribution,
tensile surface stress and interactions with single water molecules of a natively oxidised
Si surface model previously obtained by means of accurate density functional theory 
simulations.
We have applied the potential to the case of hydrophilic silicon wafer bonding at
room temperature, revealing maximum room temperature work of adhesion values for 
natively oxidised  and amorphous silica surfaces of 97~mJ/m$^2$ and 90~mJ/m$^2$,
respectively, at a water adsorption coverage of approximately 1 monolayer.
The difference arises from the stronger interaction of the natively oxidised surface 
with liquid water, resulting in a higher heat of immersion (203~mJ/m$^{2}$ vs.\
166~mJ/m$^{2}$), and may be explained in terms of the more pronounced water structuring 
close to the surface in alternating layers of larger and smaller density with respect to 
the liquid bulk.
The computed force-displacement bonding curves may be a useful input for cohesive 
zone models where both the topographic details of the surfaces and the dependence of 
the attractive force on the initial surface separation and wetting can be taken into account.
\end{abstract}

\maketitle

\section{\label{sec:1}Introduction}

Direct wafer bonding has emerged as an important technology for silicon-based
microelectronic and micromechanical systems.~\cite{Tong-book,wafer-review}
In particular, the preparation of silicon-on-insulator devices takes advantage of 
the strong adhesion between oxidised and hydrated Si surfaces to bond together 
crystalline Si wafers without the need for adhesives or high 
pressures.~\cite{lasky-exper,shimbo-exper}
The first step of the bonding process is the preparation of hydrophilic Si surfaces,
in which a $\sim$0.5~nm thin layer of native oxide covers the Si bulk and is terminated
by chemisorbed hydroxyl groups and hydrogen-bonded water molecules.
Afterwards, the surfaces are put in contact at room temperature by applying a
small pressure in a localised region of the wafers.
This triggers the propagation of a bonding front to the whole surface 
area,~\cite{Spearing-bondingfront} in a process that is driven by the 
formation of a hydrogen-bonded water network trapped between the 
Si surfaces.~\cite{Tong-book}
Finally, the bonded system is annealed at high temperature to induce 
condensation of adjacent silanols on opposite surfaces and diffusion 
of trapped water molecules away from the interface, forming a SiO$_{2}$
layer between the Si wafers.

Crucial for the formation of a defect-free bonded interface is complete bonding 
during the room temperature contacting step.
To this end, the prepared surfaces need to be sufficiently flat and clean and the 
non-covalent interactions between them sufficiently strong.
The experimental evidence points towards a dependence of the interaction strength
on the amount of water physisorbed on the surfaces,~\cite{michalske} which is 
approximately 1~monolayer (ML) at a relative humidity of about 70\,\%.~\cite{water_coverage}
Moreover, under similar humidity conditions, the adhesion strength depends substantially
on the details of the oxide structure covering the surfaces.
The measured work of adhesion for natively oxidised surfaces with 1-2~ML of adsorbed 
water is approximately 100~mJ/m$^2$.~\cite{Tong-review}
In the case of thicker oxide layers, obtained after thermal oxidation of the Si surface,
lower values in the range 60-85~mJ/m$^2$ have been reported.~\cite{maszara-thermal}
These are consistent with the bonding energy associated with the closure of cracks in 
vitreous silica glass.~\cite{michalske}
Here the measured work of adhesion is about 75~mJ/m$^2$ for relative humidities 
above 20~$\%$, but is found to rapidly decrease at lower humidity.
Notably, equal adhesion strengths could be measured in experiments of crack
opening or closing,~\cite{michalske} pointing towards a reversible adhesion mechanism 
governed by hydrogen bonds and ruling out the formation of covalent bonds between 
the surfaces at room temperature.
This is supported by a molecular dynamics simulation of amorphous silica wafers, 
which suggests a water-mediated adhesion mechanism in the absence of siloxane 
bridges across the bonding interface.~\cite{wafer}

In this work, we perform classical molecular dynamics (MD) simulations to investigate 
the crucial role of trapped water during the room temperature stage of hydrophilic
Si wafer bonding.
In particular, we aim to investigate surfaces terminated by a native 
oxide layer, whose structure and composition have been previously determined 
by means of extensive first principles molecular dynamics (FPMD) simulations 
based on density functional theory.~\cite{Lucio-oxide,Lucio-water}
For this purpose, we develop a classical force field capable of reproducing the
interactions between all chemical species present in the Si/SiO$_{x}$/water interface 
system previously obtained by FPMD.
Our starting points are the well-known Stillinger-Weber potential,~\cite{still-web} 
which has been widely used to simulate Si bulk crystal structures, and a potential 
developed by Vashishta and co-workers to study the structure and energetics of 
SiO$_{2}$ polymorphs.~\cite{vash}
Here we combine these two potentials, extending them to take into account the 
full range of Si oxidation states present in the native oxide layer as well as the 
interactions between the oxide layer and liquid water.

After describing the details of the potential development (Section~\ref{sec:2}) and the
technical details of the MD simulations (Section~\ref{sec:3}), our investigation will
start with the study of the structural and energetic details of the interfaces
of both amorphous silica and natively oxidised Si with liquid water 
(Section~\ref{sec:4a}).
Subsequently, results of our investigation into the hydrophilic bonding of 
pairs of silica wafers and natively oxidised Si wafers will be reported 
and compared (Section~\ref{sec:4b}).
Finally, all of our main findings will be briefly discussed and summarised in 
Section~\ref{sec:5}.

\section{\label{sec:2}Force Field Development}

\subsection{\label{sec:2a}Reference model structure}

Under normal atmospheric conditions, the Si surface is passivated by a hydroxylated 
native oxide layer less than 1~nm in thickness.
In Ref.~\onlinecite{Lucio-oxide}, the oxidation of the Si(001) surface in a dry 
environment has been investigated by means of FPMD simulations based on 
density functional theory (DFT).
In good agreement with medium-energy ion scattering experiments,~\cite{oxide_depth}
it was found that oxygen spontaneously adsorbs onto the bare surface up to a coverage of 
1.5~ML.
At higher coverages, further oxide growth becomes limited by diffusion of O$_2$ 
molecules to the reactive Si/SiO$_x$ interface.
In a subsequent work, water molecules were found to spontaneously adsorb and 
dissociate on the resulting oxide structure, creating further reactive sites for 
reactions with oxygen molecules.~\cite{Lucio-water}
The oxidised and partially hydrated surface structure obtained in this series
of FPMD simulations after the adsorption of seven oxygen and two water molecules
is shown in Fig.~\ref{fig:1}.
The resulting concentration of 4 hydroxyl groups on a 1.2~nm$^2$ surface is 
consistent with values ranging between 2.6 and 4.6~OH/nm$^{2}$ measured 
experimentally on amorphous silica surfaces.~\cite{Pantano_exper,Russian}
The 13 oxidised Si atoms in the final structure present variable formal valence 
charges ranging from  Si$^{+}$ to Si$^{4+}$, in ratios which are roughly 
consistent with the available experimental literature.~\cite{oxide_depth}
We have performed a Mulliken population analysis of this structure (projecting 
the ground state wave functions onto a basis consisting of s and p atomic orbitals) 
and found that the charges on the Si atoms increase approximately linearly with 
the number of nearest-neighbour O atoms, which act as electron acceptors 
(Table~\ref{t:Mulliken}).

We have shown, in a recent work, that surface oxidation in a dry environment 
is accompanied by the development of tensile surface stress up to $2.9$~N/m at 
a coverage of 1.5~ML.~\cite{b_p_paper}
The surface stress $g$ is defined as the basis invariant trace of the stress 
tensor~$\sigma$ multiplied by half the height $a$ of the periodically 
repeated supercell:~\cite{Van}
\begin{equation*}
g = \frac{1}{2}a\:\mbox{Tr}(\sigma_{\alpha\beta}) =
\frac{1}{2}a (\sigma_{yy}+\sigma_{zz}),\\
\end{equation*} 
From the above equation, using the same DFT technique as in Ref.~\onlinecite{b_p_paper}
(see Ref.~\onlinecite{DFT_details}), we have calculated the surface stress present in the 
relaxed hydroxylated native oxide, obtaining a value of 2.5~N/m (after subtracting 
the surface stress of the bottom hydrogen-terminated surface).
This value is slightly lower than for the dry oxidised structure, since the
dissociative adsorption of water on the native oxide has broken strained Si--O 
bonds, although the surface clearly remains under tensile stress.

\subsection{\label{sec:2b}Potential form for the natively oxidised Si surface}

We turn now to the development of a charge-based classical potential, whose 
analytic form and parameters will be constructed so as to reproduce the 
structure and tensile stress of the {\it ab initio} reference model described above.
Our starting points are the well-known Stillinger-Weber potential~\cite{still-web} (SW), 
which has been widely used to simulate bulk properties of silicon, and a 
potential developed by Vashishta et al. to simulate bulk and surface properties 
of silica polymorphs~\cite{vash,Trioni} (VP).
Both of these potential energy functions contain only two- and three-body particle 
interactions and may be expressed by the general form:
\begin{equation*}
V = \sum_{1\leq i < j}V_{ij}({\bf r}_{ij}) + 
\sum_{1\leq i < j < k}V_{ijk}({\bf r}_{ij},{\bf r}_{ik},{\bf r}_{jk})\,,
\end{equation*}
where the indices $i$ and $j$ run over all atoms of the simulated system and
${\bf r}_{ij}$ is the vector connecting atom $i$ with atom $j$.
Our aim here is to combine the SW and VP potentials into a single form and extend
them to include the full range of Si oxidation states present in the native oxide
layer.
In order to ensure that our potential reduces to the original forms~\cite{still-web,vash}
in the limits of bulk Si and bulk SiO$_{2}$, cut-off functions to the two- and
three-body interactions will be introduced to smoothly interpolate between potential 
regimes.
The form of these cut-off functions will be chosen so as to ensure that the potential 
energy and the forces on the particles are continuous functions of the particles' coordinates.
This is essential to account for possible breaking and forming of Si--O bonds
in MD simulations of the surface with our potential form.

In the VP potential, atomic charges with values of $+1.6$~e$^-$ and $-0.8$~e$^-$ are 
assigned to the Si$^{4+}$ and O$^{2-}$ species, respectively.~\cite{vash}
To maintain the original potential form in the case of Si$^{4+}$ species,
we assign to the oxidised Si atoms and to the O atoms  quantised partial 
charges according to:
\begin{eqnarray*}
q_{Si} = +0.4\sum_O f_c(r,R_q,\Delta_q)\\
q_{O} = -0.4\sum_{Si} f_c(r,R_q,\Delta_q) \,.
\end{eqnarray*}
According to the expression above, in agreement with the trend in computed Mulliken
charges (see Table~\ref{t:Mulliken}), the charges on the oxidised Si species
increase linearly with the number of oxygen neighbours up to a value of +1.6
for fully oxidised Si$^{4+}$ species.
The number of atomic neighbours is defined by the cut-off function $f_c$, 
which falls off smoothly and with continuous derivative over a distance 
interval $2\Delta$ around a cut-off distance $R$:
\begin{equation*}
f_c(r,R,\Delta) = \left\{ \begin{array}{ll}
1, & r<R-\Delta\\
1-\frac{r-R+\Delta}{2\Delta}+\frac{sin[\pi(r-R+\Delta)/\Delta]}{2\pi}, 
& R-\Delta\leq r<R+\Delta\\
0, & r\geq R+\Delta \, .
\end{array} \right.
\end{equation*}

As anticipated above, the Si--Si two-body interactions are as in the SW potential,
with the addition of a Coulomb term arising from the possible presence of charges on 
the Si atoms:
\begin{equation*}
V_{Si-Si'} = A(q_{Si},q_{Si'})[Br^{-4}-1]\mbox{exp}(\sigma(r-a)^{-1})
f_c(r,R_{Si},\Delta_{Si})+\frac{q_{Si}q_{Si'}}{r} \, .
\end{equation*}
In the {\it ab initio} reference structure, the Si--Si bond lengths are observed to be
approximately independent of the Si oxidation state.
Therefore, in an adjustment to the original expression,~\cite{still-web} the SW 
parameter $A$ in the equation above includes a charge dependence, 
which counteracts the Coulomb repulsion between oxidised Si atoms (Fig.~\ref{fig:2}):
\begin{equation*}
A(q,q')=A_0(1+3.2qq')f_A(q)f_A(q') \, .
\end{equation*}
Si$^{4+}$ species, as found in the tetrahedral form of SiO$_2$, form no direct 
Si--Si bonds and are instead connected via bridging oxygen atoms.
Hence, the factor $f_A(q)$ is used here to remove the Si--Si attraction for Si$^{4+}$ 
species, recovering the purely repulsive behaviour of the original VP potential: 
\begin{eqnarray*}
f_A(q)=\left\{ \begin{array}{ll}
1.0 & q \leq 1.2\\
4.0-2.5q & 1.2<q<1.6\\
0.0 & q\geq 1.6 \, .
\end{array} \right.
\end{eqnarray*}

The form of the Si--O two-body interaction is based on Ref.~\onlinecite{vash} and 
is made up of three terms: a short range steric repulsion, a charge-dipole attraction
and a Coulomb attraction:
\begin{equation*}
V_{Si-O} = \left(\frac{C_{SiO}(q_{Si})}{r^9}-\frac{D_{SiO}}{r^4}\mbox{exp}(-r/b)\right)
f_B(q_{Si}) + \frac{q_{Si}q_{O}}{r} \, ;
\end{equation*}
\begin{equation*}
C_{SiO}(q)=C_0-C_1 q \, ;
\end{equation*}
\begin{eqnarray*}
f_B(q)=\left\{ \begin{array}{ll}
2.5q & q \leq 0.4\\
1.0 & q > 0.4 \, .
\end{array} \right.
\end{eqnarray*}
In the {\it ab initio} reference structure, the Si-O bond lengths decrease 
approximately linearly with increasing oxidation state of Si.
By varying the hard-core repulsion parameter $C_{SiO}$ as a function 
of charge, we aim to reproduce the increase in bond length for oxidised 
Si atoms other than Si$^{4+}$  (Fig.~\ref{fig:3}).
The cut-off function $f_B(q)$ is introduced to reduce  the uncharged, 
sub-surface Si atoms' interactions with O atoms in the oxide layer 
in order to recover the pure SW potential form within the Si bulk.

The O--O interaction is exactly as in Ref.~\onlinecite{vash} and contains the same 
contributions as the Si--O interaction:
\begin{equation*}
V_{O-O'} = \frac{C_{OO'}}{r^7}-\frac{D_{OO'}}{r^4}\mbox{exp}(-r/b) + 
\frac{q_{O}q_{O'}}{r} \, .
\end{equation*}
The three-body interactions are all of the same SW form with parameters adapted from the 
literature,~\cite{still-web,vash,wat} again smoothed by cut-off functions $f_{c}$ to 
recover the original VP potential:
\begin{equation*}
V_{ijk} = \lambda\mbox{exp}[\gamma_1 (r_{ij}-d_1)^{-1}+\gamma_2 (r_{ik}-d_2)^{-1}]
(cos\theta_{jik}-cos\theta_0)^2f_c(r_{ij},R_{Si},\Delta_{Si})f_c(r_{ik},R_{Si},\Delta_{Si}) \, .
\end{equation*}
Finally, the hydroxyl groups that terminate the oxide surface are assigned charges equal to:
\begin{equation*}
q_{O} = -0.4\sum_{Si} f_c(r,R_q,\Delta_q) - 0.2
\end{equation*}
\begin{equation*}
q_{H} = +0.2 \, ,
\end{equation*}
consistent with the charge of approximately $-0.6$ assigned to the oxygen atoms of 
terminal silanol groups on the quartz surface in Ref.~\onlinecite{full-params}.
The O--H bond length is constrained to 0.975\AA{} and the Si--O--H interaction takes 
the above three-body form, with the parameters chosen to reproduce the variation about 
the minimum energy of the Si--O--H three-body interaction of Ref.~\onlinecite{full-params}.

It should be emphasised that the potential form described here reduces exactly 
to the original SW and VP potentials for bulk Si and SiO$_2$ under equilibrium 
conditions.
The cut-off distance $R_{Si}$ is chosen to ensure that two- and three-body 
interactions involving Si are switched on when Si--Si separations are in the region 
of 2.6 to 2.8~\AA{} or below.
In bulk Si (equilibrium bond length 2.36~\AA{}), our potential reproduces the SW 
potential up to bond lengths of 2.6~\AA{}, after which the attraction is smoothly 
reduced to zero, as shown by the dashed curve in Fig.~\ref{fig:2}.
Similarly, our potential may be applied to SiO$_2$ polymorphs (Si--Si nearest neighbour 
peak separation of 3.1\AA{} in amorphous silica, for example), reducing exactly to the 
VP potential for Si--Si separations greater than 2.8~\AA{}  (apart from an additional 
Si--Si hard-core repulsion present in the original VP form, which is negligible 
for separations above 1~\AA{}).
Below this Si--Si separation, we introduce repulsive Si--Si--O and Si--Si--Si 
three-body interactions, while the Si--Si two-body interaction is removed entirely by  
the cut-off function $f_A(q)$.

Starting from the original values of the parameters for the separate SW and VP 
potentials,~\cite{still-web,vash} the values of all parameters used in our two- 
and three-body interactions have been carefully adjusted to reproduce as accurately
as possible the {\it ab initio} reference structure in Fig.~\ref{fig:1}, and are listed in 
Table~\ref{tab:1}.
Average bond lengths and angles after minimisation of the hydroxylated 
surface structure, using this parameter set, show reasonable agreement 
with the {\it ab initio} minimised structure (Table~\ref{tab:2}).
The calculated surface stress with our potential parameters is 2.0~N/m, 
consistent with the {\it ab initio} computed tensile stress of 2.5~N/m.
Importantly, no change in the topology of the SiO$_{x}$ network (i.e. no 
breaking or forming of Si--O bonds) was observed during 50~ps of classical 
MD simulation at 300~K, during which the 
charges and potential parameters were updated at every time step.

As an additional test of our potential, we have performed a dynamical 
simulation of a Si/SiO$_{2}$ interface system composed of a 11~\AA\ thick 
slab of Si and a 17~\AA\ thick slab of $\alpha$-quartz.
At the interface, the Si(001) surface has been matched with the SiO$_{2}$ 
surface by expanding the $\alpha$-quartz lattice parameter by 10~\% in one 
direction to form an ideal, defect-free heterojunction.
The system was separately annealed for 100~ps at 500~K, 100~ps
at 700~K and for 50~ps at 1000~K.
In none of these systems was breaking of existing bonds or formation of 
new bonds observed.
Moreover, the average Si--O bond lengths and Si--O--Si angles at the interface 
have been found to decrease from their bulk values by about 1~\%  and 4~\%\, 
respectively, in agreement with a number of previous 
investigations.~\cite{sisio2_int,wat_sisio2_int}
This demonstrates a certain degree of transferability of our
potential to study Si/SiO$_{x}$ heterogeneous systems.
As a note of caution, we must say that we do not necessarily expect our potential 
to be predictive as far as the formation of unknown structures during chemical 
reactions (such as oxidation processes) is concerned.
However, this may become possible upon augmentation of the potential 
within hybrid quantum/classical schemes such as the recently developed 
``Learn on the Fly'' technique.~\cite{LOTF04}

\subsection{\label{sec:2c}Interactions between the oxidised surface and water}

To study the behaviour of SiO$_{x}$ surfaces in a wet environment,
we model the interactions between the surface atoms and water
molecules as a sum of Coulomb (via atomic partial charges) and 
Van der Waals (VdW) contributions.
Consistent with force fields used to simulate solvated biomolecules
in solution, the VdW interactions between water molecules and the 
surface hydroxyl groups are described by a  ``hydrogen bond''  form 
$V(r_{ij})=A/r_{ij}^{12}-B/r_{ij}^{10}$, with the $A$ and $B$
parameters taken from standard AMBER parameter sets.~\cite{amber}
The VdW interactions between water molecules and all other 
atom types are of the standard Lennard-Jones (LJ) form:
\begin{equation*}
V(r_{ij})=4\epsilon_{ij}\left(\left(\sigma_{ij}/r_{ij}\right)^{12}-
\left(\sigma_{ij}/r_{ij}\right)^{6}\right) \, .
\end{equation*}
For each pair of atoms we define:  $\epsilon_{ij}=\sqrt{\epsilon_i\epsilon_j}$ 
and  $\sigma_{ij}=2^{-1/6}(\sigma_i+\sigma_j)$, such that the minimum energy 
separation, $r_0$, is $\sigma_i+\sigma_j$,  and $V(r_0)=-\sqrt{\epsilon_i\epsilon_j}$.

It has to be noted that the values of the partial charges on the surface atoms 
have been chosen so as to guarantee stability of the SiO$_{x}$ network (see 
previous section).
In particular, their values differ from the atomic point charges which 
best fit the electrostatic potential in a region outside the VdW radius 
of the atoms (ESP charges), which would be the best choice of charges
to simulate the surface hydration properties.~\cite{ESP}
For this reason, we have adapted the parameters of the LJ part of the non-bonded
surface-water potential so as to reproduce the binding energy curves for 
isolated water molecules on selected surface sites obtained in {\it ab initio} calculations.
This energy is mostly accounted for by electrostatic effects, which are well described 
within standard DFT techniques.

Within DFT,~\cite{DFT_details} we have calculated the total energy of a single 
water molecule as a function of its distance from the hydroxylated surface for two
configurations, chosen such that the main interactions present are between one H atom
of water with an O atom of the surface and between the
O atom of water with a Si$^{4+}$ atom of the surface (Fig.~\ref{fig:4}).
Identical total energy calculations were performed using the TIP3P water 
model~\cite{tip3p} and our newly developed force field for the native 
oxide, for different choices of the LJ parameters of the water-surface 
interactions.
The best obtained binding energy curves are shown in Fig.~\ref{fig:4},
and the corresponding, optimised set of LJ parameters is reported in 
Table~\ref{tab:3}.
As will be shown in Section~\ref{sec:4a}, with this set of parameters the 
calculated heat of immersion of an amorphous silica surface amounts
to 166~mJ/m$^{2}$, which should be compared with the experimental
value of 158 mJ/m$^{2}$, measured for a surface with a density of
terminal --OH groups of 3.4 nm$^{-2}$.~\cite{H_imm2}

\section{\label{sec:3}Computational Details}

All classical simulations were performed using the  DL\_POLY\_3 
MD package,~\cite{dlpoly3} in which our newly-developed force field 
for the native oxide could be easily implemented.
Long-ranged Coulomb interactions were treated using the Ewald 
sum, with a real space cut-off of 8~\AA{}.
Newton's equations of motion were integrated with the velocity 
Verlet algorithm and a time step of 1~fs, using the RATTLE 
algorithm~\cite{rattle} to constrain O--H bond lengths.
Where necessary, temperature control was achieved by velocity rescaling.

An amorphous silica model was generated according to existing
MD schemes,~\cite{Garowet} using the VP potential in its original
form,~\cite{vash} which has been successfully applied to study 
amorphous silica surfaces.~\cite{Trioni}
Namely, we initially placed 1314 randomly positioned Si and O atoms, 
in the ratio $1:2$, in a  periodically repeated 
$30\!\times\!30\!\times\!20$~\AA$^{3}$ simulation cell.
The initial structure was annealed for 100~ps at 8000~K, then 
for a further 100~ps at 4000~K.
Subsequently, the system was cooled to 300~K over a period of 360~ps, 
switching to a constant pressure ensemble controlled by a Nos\'{e}-Hoover 
combined thermostat and barostat with 0.2~ps relaxation times.
Following a further 200~ps annealing at 300~K, the bulk density was 
2.37~g/cm$^{3}$ (compared to the experimental value of 2.20~g/cm$^{3}$).~\cite{vash}
Further structural analysis, detailed in Fig.~\ref{fig:rdf} and Table~\ref{tab:rdf}, 
was in good agreement with the literature~\cite{vash} and revealed the presence 
of a network of corner-sharing SiO$_2$ tetrahedra.

Starting from the bulk amorphous silica structure obtained, a 50~\AA{} vacuum 
layer was inserted in the $z$-direction of the simulation cell, resulting in separated, 
periodically repeated slabs of surface area 9.1~nm$^2$.
The resulting surfaces were annealed at 1000~K for 500~ps, cooled to 300~K over 
750~ps and, finally, annealed at 300~K for a further 500~ps, all at constant 
volume.
These long annealing times were necessary to stabilise the concentrations of 
defects, found here in the form of threefold coordinated Si atoms 
and dangling O atoms, with converged concentrations of 1.6~nm$^{-2}$
and 1.7~nm$^{-2}$, respectively.
These concentrations may be compared to Ref.~\onlinecite{Garofalini}, which uses a 
Born-Mayer-Huggins pair interaction and a similar three-body potential, obtaining 
concentrations of the two defect types of 0.6~nm$^{-2}$ and 1.9~nm$^{-2}$.
Fig.~\ref{fig:surf_defects}, left, reveals that these defects were located in 
the first 5~\AA{} of the surface.
In agreement with an existing VP potential simulation of the amorphous silica 
surface,~\cite{Trioni} we found that the surface was terminated by dangling O atoms and that 
surface O--Si--O bond angles were shifted to higher angles, indicating the presence 
of under-coordinated Si, as shown in Fig.~\ref{fig:surf_defects}, right.
After terminating these two defect types with --OH and --H groups, respectively,
and relaxing the whole system, the resulting hydroxylated amorphous silica 
surface had a hydroxyl group surface concentration of 3.3~nm$^{-2}$,
which lies in the range of experimental measurements 
(2.6 - 4.6~nm$^{-2}$).~\cite{Pantano_exper,Russian}

To model the hydroxylated native oxide on Si(001), the {\it ab initio} 
reference structure shown in Fig.~\ref{fig:1} was periodically 
repeated in the surface plane to form a $3\!\times\!3$ slab of 
surface area 10.6~nm$^2$.
The bottom surface was terminated with a copy of the oxide layer
after rotation through 180$^\circ$ and translation by a bulk Si 
lattice parameter.
The resulting slab consisted of 792 Si atoms, corresponding to eleven layers, 
oxidised and hydroxylated on either side with a total of 288 O and 72 H atoms.
The lattice parameter was fixed to the equilibrium value for bulk Si  obtained
with the SW potential (5.44~\AA), and the vacuum gap in the 
direction perpendicular to the surface plane was set to 50~\AA{}, as in the
case of the amorphous silica slab.

To investigate water layering at the two hydroxylated surfaces (Section~\ref{sec:4a}), 
the 50~\AA{} vacuum layers were filled with TIP3P water molecules~\cite{tip3p} 
at a density of approximately 1~g/cm$^{3}$.
For each surface, the entire system was equilibrated at 300~K and the height 
of the supercell adjusted to remove the stress in the direction perpendicular 
to the surface.
Particle coordinates were collected every 100 time steps over a 1~ns production run.
Selected snapshots from these simulations were used to construct the initial input files
for the simulations of hydrophilic wafer bonding (Section~\ref{sec:4b}), keeping in the simulation
cells only the water molecules closest to the surfaces up to heights corresponding to
the desired surface water coverages.
Based on the structural analysis of the surface/water interfaces (Section~\ref{sec:4a}),
one ML of adsorbed water is found to contain 98 and 117 water molecules on the 
amorphous silica and native oxide surface models, respectively, which correspond
to surface densities of 10.7 and 11.0 molecules/nm$^{2}$.

Based on Ref.~\onlinecite{wafer}, force-displacement curves were calculated by 
reducing the height of the simulation cell perpendicular to the surfaces (and thus 
the separation between the top surface and the periodic image of the bottom surface) 
at a rate of 0.1~\AA{} every 11~ps.
At each separation distance, the system was first equilibrated and thermalised
to 300 K by velocity rescaling for 1~ps.
During the subsequent 10~ps, the particle coordinates were collected every 0.2~ps and the 
average $z$ component of the stress in the simulation cell was calculated.
In this way, the net interfacial force could be
computed as a function of separation distance and the resulting curve integrated 
to obtain the total surface bonding energy.

\section{\label{sec:4}Results and Discussion}

\subsection{\label{sec:4a}Structure and energetics of the SiO$_{x}$/Water interface}

We start our investigation of the interactions between wet oxidised silicon surfaces
by studying the structure of bulk water in contact with both the amorphous
silica and the natively oxidised Si surfaces.
Any surface in contact with bulk water will have an effect on the intrinsic ordering
of water in its proximity.
In particular, water molecules will interact strongly with and partly penetrate into
hydrophilic surfaces to form surface-water hydrogen bonds, in competition with
water-water hydrogen bonds in the liquid.
This will result in a layered structure of water molecules close to the surface,
which can be quantified by the surface density profile of water molecules 
along the direction perpendicular to the surface. 

We have calculated the density profiles of water in contact with the two
surface models by computing the average number of water
molecules present in 0.1~\AA{} thin, planar slices parallel to the surface
during a 1~ns simulation, as described in Section~\ref{sec:3}.
The results are reported in Fig.~\ref{fig:5}.
In each case, we found evident structuring of water in layers of higher and lower
density with respect to the bulk, consistent with previous simulations of a 
hydrophilic quartz surface.~\cite{full-params}
For both surfaces, the position of the main peak in the density profile matches 
the position of the outermost surface hydroxyl group.
The presence of subsidiary peaks closer to the surface, especially in the case of 
natively oxidised Si, is indicative 
of water penetration into the relatively open structure of the hydroxylated
surfaces, as expected.
With respect to the position of the first maximum (of density 1.05~g/cm$^{3}$), 
the water density close to the amorphous silica surface presents a trough at a 
distance of 1.1~\AA{} and a second maximum at 2.3~\AA{}, before reaching the 
bulk water density of 0.98~g/cm$^{3}$ at a distance of approximately 4~\AA{}.
The main peak at the native oxide surface is higher, reaching 1.32~g/cm$^{3}$, 
and the oscillations continue out to approximately 5~\AA{} from the position
of the main peak before reaching the bulk water density.

The shape of the density profile obtained allows us to define a~ML 
as the layer of water molecules contained between the surface and the position 
of the first trough after the main density peak.
By integration of the density profiles, 98 and 117 water molecules are found to
be contained in 1~ML on the amorphous silica and the natively oxidised surfaces,
corresponding to surface densities of 10.7 and 11.0 water molecules/nm$^{2}$,
or to surface areas per water molecule of 9.3 and 9.1~\AA$^{2}$, respectively.
These values should be compared with early assumptions,~\cite{water_coverage}
where an area of 10.6~\AA$^{2}$ has been assigned to a water molecule adsorbed
on quartz, the difference arising probably from the number of molecules which
are able to penetrate relatively deeply into the surface oxide structures.

For each surface, we have computed the heat of immersion as the difference 
between the energy of the system in contact with water and
the energies of the two separate components, namely the dry 
surface model and bulk water. 
For each immersed system, the energy has been computed as the average value of
the potential energy obtained in separate MD simulations at 300~K for 1~ns.
With this approach, we calculated heat of immersion values of 
166~mJ/m$^{2}$ and 203~mJ/m$^{2}$ for the amorphous silica 
and native oxide respectively.
The first of the two values may be compared with experimentally
measured values of 157~mJ/m$^{2}$ or 158~mJ/m$^{2}$,
for surfaces with estimated densities of terminal --OH groups 
of 2.3~nm$^{-2}$ or 3.4~nm$^{-2}$.~\cite{H_imm1,H_imm2}
The excellent agreement between the experimental and the theoretical
values provides good support for the parametrisation of our 
surface-water potential, as described in Section~\ref{sec:2c}.
We note that a possible deprotonation of the surface hydroxyl groups
in the experimental system is not expected to influence substantially 
the results obtained.
In fact, in deionised water at pH~7, the surface charge is expected to 
be of the order of -1~$\mu$C/cm$^{2}$,~\cite{surf_charge} corresponding 
to just one deprotonated hydroxyl group  per simulation cell, which is 
unlikely to contribute significantly to the computed heat 
of immersion.

A further insight into the details of the surface/water interfaces may be
gained by analysing the average number of surface-water and 
water-water hydrogen bonds formed per water molecule within the
first ML compared with the liquid bulk.
Considering a hydrogen bond to be present between two O atoms when the 
O--H$\cdots$O angle is greater than $140^{\circ}$ and the 
O--O separation is smaller than 3.5~\AA{}, the calculated average number
of hydrogen bonds per molecule in bulk water is 3.13.
Close to the surface, within the first ML, the corresponding values
for the amorphous silica and the native oxide are 3.30 and 3.27, respectively,
reflecting the competitive hydrogen bond formation at the surface.
In each case, 0.76 hydrogen bonds per molecule are formed
with the surface (Table~\ref{tab:HB}), which can be further divided
between the bonds donated to O atoms bridging Si atoms of the surface,
the bonds donated to O atoms of terminal hydroxyl groups, and the bonds
received by H atoms of hydroxyl groups.
The calculated values (Table~\ref{tab:HB}) indicate evident differences
between the two surfaces.
Namely, more hydrogen bonds are formed between water and bridging
oxygen atoms in the case of the native oxide structure, probably due to the
reduced charge on some of the Si atoms, and thus the reduced electrostatic 
screening, compared with the amorphous silica surface.
Since the charges on the bridging oxygen atoms are higher than those on the 
hydroxyl groups, this may help to explain the difference between the
computed heat of immersion in the two cases, despite the small difference
in the number of water molecules per surface area within the first ML.

\subsection{\label{sec:4b}Wafer Bonding}

In the previous subsection, we have shown how competitive hydrogen 
bond formation at the surface/water interface leads to a local restructuring
of the hydrogen bond network between the molecules.
This results in oscillations in the water density, which propagate several 
Angstroms into the bulk liquid. 
The thin layer of water trapped between oxidised silicon wafers in
the room temperature stage of hydrophilic wafer bonding will be
subjected to the same effect, whose relationship to the attractive
force between the wafers will be investigated below.

We have calculated the average net force present between two
bare amorphous silica surfaces and two bare native oxide surfaces 
in the absence of any water molecules as a function of the interface
separation according to the procedure described in Section~\ref{sec:3}.
Starting from two separated surfaces, no attraction is observed
when the surfaces are moved together  (dashed lines in Fig.~\ref{fig:6}, top).
When the surfaces become very close to each other, steric repulsion
occurs between the surface atoms and the repulsive force increases rapidly.
Zero separation in the curves in Fig.~\ref{fig:6} was defined as the 
point of onset of this repulsive force.

The situation is very different in the presence of water molecules
between the surfaces.
In Fig.~\ref{fig:6}, top, we have plotted the force-separation curves
calculated at increasing water coverages (from approximately 0.25 to 
2.5~ML per surface), along with the associated hydrogen bond density 
as a function of surface separation (Fig.~\ref{fig:6}, bottom).
As the surfaces move together, they experience a net attractive force
(negative values in the reported plots), which initially increases, reaches
a maximum value and then decreases, becoming repulsive at 
small separations.
Subsequent debonding simulations, in which the surfaces were pulled apart 
at the same rate, revealed no hysteresis in the force-separation curves.

The overall behaviour of the system may be rationalised by looking at the
number of hydrogen bonds formed per molecule between the surfaces.
Starting from larger separations, the hydrogen bond density increases as the surfaces 
come into closer and closer contact.
At smaller separations, if the concentration of water is smaller than
about 1~ML (dotted lines in Fig.~\ref{fig:6}), the interaction between
the surfaces becomes repulsive well before the hydrogen bond density reaches the 
equilibrium values of 3.30 or 3.27 at the amorphous silica or native 
oxide surfaces, respectively (see Section~\ref{sec:4a}).
In these cases, the computed force becomes dominated by the repulsion 
between the two solid surfaces, and the repulsive part of the force-separation 
curve tends to that calculated in the absence of trapped water.
On the other hand, for water coverages greater than approximately 1~ML 
(solid lines in Fig.~\ref{fig:6}), the onset of repulsion occurs before the 
surfaces interact directly.
In fact, as is visible in Fig.~\ref{fig:6}, the point of zero force between
the surfaces at increasing water coverages tends to the point at which the 
density of hydrogen bonds reaches 3.13, the equilibrium value in bulk water.

These results imply that there is an optimum water concentration for room 
temperature hydrophilic wafer bonding, low enough that there is a high 
concentration of energetically favourable surface-water hydrogen bonds, 
as observed within the first ML of the systems in Section~\ref{sec:4a}, but high enough 
that the oxide surface repulsion does not dominate.
By integration of the force-displacement curves we can compute the
energy gained during the simulated bonding process, that is the work of
adhesion of the wet surfaces $Q_{ad}$.
This is reported as a function of the water coverage in Fig.~\ref{fig:7}, 
indeed showing an optimum water concentration for bonding at approximately 
1~ML for both surface models.
At very low coverages, the steric repulsion between the surface dominates,
so that $Q_{ad}$ rapidly decreases to negligible values.
At the other extreme, for high coverages, the values of $Q_{ad}$
are very similar in the two cases and tend to the value of bulk TIP3P water,
calculated by integration of a force-separation curve for a water-water 
interface, in the absence of any solid surface.
The value obtained, 78~mJ/m$^{2}$, is indicated with a dotted line
in Fig.~\ref{fig:7} and may be compared with the experimental surface 
energy of 72~mJ/m$^2$ for the water-air interface at room 
temperature.~\cite{water-tens}

For the case of amorphous silica, the variation of $Q_{ad}$ with 
water coverage may be compared with forces associated with crack closure
in vitreous silica glass at different relative humidities, as reported 
in Ref.~\onlinecite{michalske}.
In this comparison, we shall take into account that $Q_{ad}$ is
half the strain energy release rate, and that each value of relative
humidity is associated with a well-defined coverage of adsorbed water.~\cite{water_coverage}
The investigation in Ref.~\onlinecite{michalske} shows that, at high humidity, $Q_{ad}$  
is roughly constant at about 75~mJ/m$^2$, which corresponds to the surface
energy of bulk water.
At relative humidities lower than about 20~$\%$, which corresponds to a water
coverage of approximately 0.5~ML, $Q_{ad}$ starts to decrease and drops 
to  25~mJ/m$^2$ at a coverage of approximately 0.25~ML.
This in good agreement with the shape of the amorphous silica curve shown in Fig.~\ref{fig:7},
where the computed work of adhesion is 19~mJ/m$^2$ at 0.25~ML coverage,
and is comparable with the surface energy of bulk water at large coverages.
In our simulations, $Q_{ad}$ presents a small peak of 90~mJ/m$^2$
at approximately 1~ML.
Experimental values of work of adhesion in direct surface bonding experiments
using silica glasses are lower, in the range of 60 to 85~mJ/m$^2$,~\cite{maszara-thermal} 
which might be explained by the microroughness present in the experimental wafer 
samples.~\cite{Spearing_roughness}

As far as the natively oxidised Si surface is concerned, our simulations indicate
a slightly higher maximum value of $Q_{ad}$ than amorphous silica, which may
be related to the higher heat of immersion and to the more pronounced
oscillations of the density close to the surface, as reported in Fig.~\ref{fig:5}.
In fact, the water density profiles between closely spaced native oxide surfaces 
show oscillations that match those observed at an isolated slab (Fig.~\ref{fig:8}).
At 0.5~ML coverage, the main peaks from opposite surfaces overlap, producing a 
single peak in the water density at equilibrium (Fig.~\ref{fig:8}, top right),
fully consistent with trapping 1 ML of water between the two surfaces.
In this case, as shown in Fig.~\ref{fig:8}, top left, the initial force required to 
separate the surface and thus break the hydrogen bond network of this ML
is high, although $Q_{ad}$ is relatively low, due to the surface-surface repulsion 
at small separation.
At 1.0~ML coverage, Q$_{ad}$ is maximum, reaching 97~mJ/m$^2$, in 
good agreement with experimental values of approximately 100~mJ/m$^2$ for 
natively oxidised Si wafers covered by 1-2~ML of 
water.~\cite{Tong-review}
In this case, the main peaks in the equilibrium water density at the two 
surfaces are separated by 2.5~\AA{} (Fig.~\ref{fig:8}, bottom left), close 
to the ideal O--O separation of 2.8~\AA{} for hydrogen-bonded TIP3P 
water.~\cite{tip3p}
This indicates the existence of an ordered, energetically stable, hydrogen-bonded 
water network, spanning the two bonded wafers.
Finally, at higher coverages, the density in the central region between 
the wafer is constant, roughly at the equilibrium density of bulk water 
(Fig.~\ref{fig:8}, bottom right), consistent with Q$_{ad}$ being 
comparable with the surface energy of liquid water.

In an ideal debonding experiment, $Q_{ad}$ may be thought of as being composed 
of a large energy loss (positive values) associated with breaking the network of 
hydrogen bonds between the wafers and with an energy gain associated with the
rearrangement of water at the surfaces, to optimise the number of hydrogen
bonds per molecule.
The more structured the water layer, the smaller will be the energy gain,
due to the reduced capability of water to rearrange within the layer
(which is consistent with the measured lower entropy of water molecules 
at low coverages on oxidised surfaces~\cite{water_coverage}).
Therefore, the slightly larger $Q_{ad}$ value than in the case of amorphous silica 
may be explained by the more pronounced structuring of water between 
the native oxide surfaces induced by the stronger surface-water interactions 
(see Section~\ref{sec:4a}).

\section{\label{sec:5}Summary and Conclusions}

In conclusion, we have developed a force field which enables us to investigate,
at the classical level, the interactions between oxidised Si surfaces and a liquid
environment.
This is of particular importance given the increasingly wider fields of application
of silicon-based microelectromechanical devices put in direct contact with water
solution, such as sensors or actuators used in a physiological environment.~\cite{biomems}
In our potential form, the surface interacts with the external environment through 
non-bonded interactions of the electrostatic and Van der Waals type, which makes the 
combination of the developed potential with standard force fields used to 
investigate biological macromolecules straightforward, and opens the way to large-scale 
simulations of biomolecule adsorption on realistic Si surfaces models.
Moreover, the form of our force field allows processes associated with a rearrangement 
of the Si--O bond network to be simulated, as is necessary for the application of classical 
Hamiltonians in hybrid quantum-classical schemes 
of the ``Learn on the Fly'' type.~\cite{LOTF04}
These appear to be very promising tools for the atomistic simulations of chemo-mechanical
effects (for example, those occurring at a crack tip during a stress-corrosion process), 
which require both large system sizes and accurate quantum description of limited portions 
of the simulated systems.~\cite{James_crack,Gian_proceeding_smartcut,Noam-friction}

In this work, we have applied the newly developed potential to the simpler case of
hydrophilic Si wafer bonding at room temperature, where no
breaking or forming of covalent bonds is expected to play a role.
We have found that our parameterisation of the interatomic potential is able to 
reproduce the structural and mechanical
properties of the natively oxidised Si surface, as obtained in previous FPMD 
simulations based on DFT.~\cite{Lucio-oxide,Lucio-water}
Moreover, calculated energetic details of the amorphous silica/water interface, 
such as the surface heat of immersion, were found to agree well with existing experimental
measurements.
As compared with the amorphous silica surface, the natively oxidised Si surface
was found to interact more strongly with the liquid water, resulting in a higher heat
of immersion (203~mJ/m$^{2}$ vs. 166~mJ/m$^{2}$) and a more pronounced
structuring of the water molecules close to the surface in alternating layers of larger
and smaller density with respect to the liquid bulk.

The layering of water at the oxidised surface has a profound effect on the computed
force-displacement curves in simulated wafer bonding experiments at different coverages
of adsorbed water. 
We have predicted that there is an optimum surface water concentration for bonding 
between both glassy silica and natively oxidised Si surfaces, for which the maximum
work of adhesion values are 90~mJ/m$^{2}$ and 97~mJ/m$^{2}$, respectively, 
occurring at approximately 1~ML coverage (corresponding to about 70~$\%$ 
relative humidity~\cite{water_coverage}).
It must be noted that these values refer to perfectly flat surfaces, whereas 
experimentally measured values are expected to depend heavily on the nanoscale
surface roughness present in the samples.~\cite{Spearing_roughness}
This is especially true in the case of bonding experiments, where a work of 
adhesion as low as 40~mJ/m$^{2}$ has been measured for two natively oxidised Si 
surfaces.~\cite{Spearing-bondingfront}
The relatively slow diffusion of water parallel to the surface will prevent 
a perfect wetting of the whole surface during bonding.
However, in subsequent fracture tests the work of adhesion has been shown to rise 
to 120~mJ/m$^{2}$, possibly due to rearrangement of the water layer over the 
length scale of the surface roughness.~\cite{Turner-submitted}
Our calculated force-displacement curves may provide a useful input for 
macroscopic models where both the topographic details of the  bonded surfaces and the 
precise dependence of the attractive force on the initial surface separation and wetting
can be taken into account.
Combining the results presented here with these macroscopic models, such as 
the cohesive zone model of Kubair and Spearing,~\cite{Spearing-czm} 
will be the subject of future work.~\cite{multiscale-wafer}

\subsection*{Acknowledgements}

We would like to thank Yohsitaka Umeno for useful discussions and suggestions.
Computational resources were provided by the Cambridge HPC Service, U.K., by the 
Zentrum f\"ur Informationsdienste und Hochleistungsrechnen, Dresden, and by the
HLRS Stuttgart within the AQUOXSIM project and through the HPC-Europa 
project (RII3-CT-2003-506079, with the support of the  European Community - 
Research Infrastructure Action of the FP6).
LCC acknowledges support by the Alexander von Humboldt Stiftung, by the Japanese
Society for the Promotion of Science and by the Deutschen Forschungsgemeinschaft 
within the Emmy Noether Programme.
This work has been supported by the EPSRC, U.K.
\bibliography{wafer_paper_01}

\clearpage

\subsection*{Tables}

\begin{table}[h!]
\caption{Distribution of Mulliken charges in the reference structure of an oxidised
               and hydrated Si surface obtained in a series of FPMD simulations. 
Charges are multiples of the electronic charge.~\label{t:Mulliken}}
\begin{center}
\begin{tabular}{l|ccccc|c}
\hline \hline
Species & \multicolumn{5}{| c |}{Mulliken Charges}  & Average  \\
\hline
Si$^{+}$         &  0.62  &  0.65  &  0.46  &  0.47  & 0.61  &  0.56  \\
Si$^{2+}$  &  1.26  &  1.28  &  1.28  &            &&  1.27  \\
Si$^{3+}$  &  1.79  &  1.82  &  1.79  &            &&  1.80  \\
Si$^{4+}$  &  2.30   &  2.34  &         &             &&  2.32  \\
\hline \hline
\end{tabular}
\end{center}
\end{table}               
          
\begin{table}[h!]
\caption{Parameters of the interatomic potential. Units of length are \AA{} and those 
of energy are e$^2/\mbox{\AA{}}(=14.39$~eV).~\label{tab:1}}
\begin{center}
\begin{tabular}{cccccccccc}
\multicolumn{8}{l}{Two-body potential parameters}\\[1mm]
\hline
\hline
\vphantom{\Large A}$R_q$        &2.0  &$A_0$    &1.062   &$C_0$     &14.871     &$C_{OO'}$     &51.692\\
                   $\Delta_q$   &0.1  &$B$      &11.603  &$C_1$     &2.178      &$D_{OO'}$     &1.536\\
                   $R_{Si}$     &2.7  &$a$      &3.771   &$b$       &4.43       &&\\
                   $\Delta_{Si}$&0.1  &$\sigma$ &2.0951  &$D_{SiO}$ &3.072      &&\\[2mm]
\multicolumn{10}{l}{Three-body potential parameters}\\[1mm]
\hline
\hline
\vphantom{\Large A}$\lambda_{SiSiSi}$&3.164&$\lambda_{SiSiO}$&3.164&$\lambda_{SiOSi}$&1.400&$\lambda_{OSiO}$&0.350&$\lambda_{SiOH}$&0.521\\
$\gamma_{1,SiSiSi}$&2.51&$\gamma_{1,SiSiO}$&4.06&$\gamma_{1,SiOSi}$&1.00&$\gamma_{1,OSiO}$&1.00&$\gamma_{1,SiOH}$&1.00\\
$\gamma_{2,SiSiSi}$&2.51&$\gamma_{2,SiSiO}$&0.52&$\gamma_{2,SiOSi}$&1.00&$\gamma_{2,OSiO}$&1.00&$\gamma_{2,SiOH}$&1.00\\
$d_{1,SiSiSi}$&3.771&$d_{1,SiSiO}$&3.981&$d_{1,SiOSi}$&2.60&$d_{1,OSiO}$&2.60&$d_{1,SiOH}$&2.60\\
$d_{2,SiSiSi}$&3.771&$d_{2,SiSiO}$&2.933&$d_{2,SiOSi}$&2.60&$d_{2,OSiO}$&2.60&$d_{2,SiOH}$&2.60\\
$\theta_{0,SiSiSi}$&109.47$^{\circ}$&$\theta_{0,SiSiO}$&109.47$^{\circ}$&$\theta_{0,SiOSi}$&141.00$^{\circ}$&$\theta_{0,OSiO}$&109.47$^{\circ}$&$\theta_{0,SiOH}$&122.50$^{\circ}$\\
\end{tabular}
\end{center}
\end{table}

\begin{table}[h!]
\caption{Comparison between {\it ab initio} and classical minimisations of the 
hydroxylated Si(001) surface. In the {\it ab initio} structure, Si--O bond lengths (\AA{}) 
increase with decreasing Si oxidation state, while Si--Si bond lengths remain 
approximately constant. In the classical simulation, although some bond lengths 
and angles have changed, there is no restructuring of the surface. Si--O--H 
angles (126$^\circ$) also agree with DFT values (127$^\circ$).~\label{tab:2}}
\begin{center}
\begin{tabular}{l|ccccc|ccccc}
\hline \hline
&\multicolumn{5}{c|}{{\it ab initio} structure}&\multicolumn{5}{c}{classical structure}\\[1mm]
&Si-O&Si-Si&O-Si-O&Si-Si-O&Si-Si-Si&Si-O&Si-Si&O-Si-O&Si-Si-O&Si-Si-Si\\[1mm]
\hline
\vphantom{\Large A}Si$^{4+}$&1.64&--&109$^\circ$&--&--&1.63&--&109$^\circ$&--&--\\[1mm]
\vphantom{\Large A}Si$^{3+}$&1.65&2.36&109$^\circ$&109$^\circ$&--&1.61&2.41&119$^\circ$&96$^\circ$&--\\[1mm]
\vphantom{\Large A}Si$^{2+}$&1.67&2.39&109$^\circ$&108$^\circ$&109$^\circ$&1.62&2.44&139$^\circ$&101$^\circ$&110$^\circ$\\[1mm]
\vphantom{\Large A}Si$^{+}$&1.70&2.38&--&108$^\circ$&111$^\circ$&1.65&2.39&--&108$^\circ$&111$^\circ$\\[1mm]
\hline \hline
\end{tabular}
\end{center}
\end{table}

\begin{table}[h!]
\caption{LJ parameters optimised for the surface-water interactions. Water OW 
parameters are taken from the TIP3P water model. Parameters for water HW atoms are for 
interactions with Si and OB only, otherwise $\epsilon_i$ is zero. In the surface, 
OB refers to bridging O atoms and OH to hydroxyl O atoms.~\label{tab:3}}
\begin{center}
\begin{tabular}{l|cc|cccc}
\hline \hline
\multicolumn{1}{c|}{}&\multicolumn{2}{c|}{water}&\multicolumn{4}{c}{surface}\\
&OW&HW&Si&OB&OH&H\\[1mm]
\hline
\vphantom{\Large A}$\epsilon_i$/eV$\times10^{-2}$  & 0.66  & 0.13  & 1.30  & 1.13  & 0.65  &0.09  \\[1mm]
\vphantom{\Large A}$\sigma_i$/\AA{}                & 1.768 & 0.650 & 1.600 & 1.762 & 1.650 & 1.000\\[1mm]
\hline \hline
\end{tabular}
\end{center}
\end{table}

\begin{table}[h!]
\caption{MD results for amorphous silica at 300K. The peak bond length or angle and 
the corresponding full width at half maximum (FWHM) of each distribution is reported 
and compared to Ref.~\onlinecite{vash}.~\label{tab:rdf}}
\begin{center}
\begin{tabular}{l|cc|cc}
\hline \hline
\multicolumn{1}{c|}{}&\multicolumn{2}{c|}{this work}&\multicolumn{2}{c}{Ref.~\onlinecite{vash}}\\
&bond length/\AA{}&FWHM/\AA{}&bond length/\AA{}&FWHM/\AA{}\\[1mm]
\hline
\vphantom{\Large A}Si--O  & 1.62  & 0.07  & 1.62  & 0.05\\
\vphantom{\Large A}O--O   & 2.63  & 0.20  & 2.64  & 0.15\\
\vphantom{\Large A}Si--Si & 3.09  & 0.27  & 3.10  & 0.20\\
\hline
&bond angle&FWHM&bond angle&FWHM\\[1mm]
\hline
\vphantom{\Large A}O--Si--O & 108$^\circ$  & 12$^\circ$  & 109.6$^\circ$  & 10$^\circ$\\
\vphantom{\Large A}Si--O--Si & 141$^\circ$  & 26$^\circ$  & 142.0$^\circ$  & 25$^\circ$\\
\hline \hline
\end{tabular}
\end{center}
\end{table}

\begin{table}[h!]
\caption{Statistics of hydrogen-bonding between the two surfaces and the
first ML of water (within the first peak of the density distribution).
\label{tab:HB}}
\begin{center}
\begin{tabular}{l | c| c c c| c }
\hline \hline
Surface &  OW--HW$\cdots$OW  &  OW--HW$\cdots$OB  &  OW--HW$\cdots$OH  &  OH--H$\cdots$OW  &  Total surface \\
\hline
a-SiO$2$  &  2.54  &  0.14  &  0.30  &  0.32  &   0.76  \\
native      &  2.51  &  0.24  &  0.26  &  0.26  &   0.76 \\
bulk water & 3.13   & --   & --  & --  & --  \\
\hline \hline
\end{tabular}
\end{center}
\end{table}

\clearpage

\subsection*{Figure Captions}

\begin{figure}[h!]
\caption{Side and top views of the hydroxylated native oxide layer on  Si(001) as 
obtained in a series of FPMD simulations.~\cite{Lucio-oxide,Lucio-water}
O atoms (red) either bridge two Si atoms or form part of a terminating 
hydroxyl group. The valence states of the 13 oxidised Si atoms range from Si$^{+}$ 
to Si$^{4+}$.
\label{fig:1}}
\end{figure}

\begin{figure}[h!]
\caption{Examples of the Si--Si two-body interaction for Si$^{0+}$--Si$^{0+}$ 
species (dashed line) and Si$^{2+}$--Si$^{1+}$ species (dotted line), compared with 
the original SW interaction for uncharged Si (solid line). The 
equilibrium bond lengths are similar and both potentials tend to the Coulomb 
interaction for large separations.}
\label{fig:2}
\end{figure}

\begin{figure}[h!]
\caption{The Si--O two-body interaction. The equilibrium bond length increases 
with decreasing Si partial charge.
\label{fig:3}}
\end{figure}

\begin{figure}[h!]
\caption{DFT (dashed lines) and classical (solid lines) binding energy 
curves of a single water molecule above the hydroxylated native oxide surface. 
In each simulation, the water molecule is displaced vertically, and the 
relative total energy is plotted as a function of its separation, d, from the 
surface. In the first configuration (left, black curves), the water molecule, with 
a H atom pointing downwards is bound to a bridging O atom. In the second 
configuration (right, blue curves), the water O atom is pointing downwards 
towards a Si atom and is repelled from the surface.
\label{fig:4}}
\end{figure}

\begin{figure}[h!]
\caption{(left) Total radial distribution function in bulk amorphous silica. The 
first three peaks correspond to Si--O, O--O and Si--Si nearest neighbours. (right) 
O--Si--O and Si--O--Si bond angle distributions, which are consistent with a 
network of corner-sharing SiO$_2$ tetrahedra.
\label{fig:rdf}}
\end{figure}

\begin{figure}[h!]
\caption{(left) Relative densities of fourfold coordinated (Si-4) and threefold 
coordinated (Si-3) Si atoms and bridging (O-2) and dangling (O-1) O atoms with 
increasing distance from the centre of the amorphous silica slab. Defects are 
concentrated in the first 5~\AA{} of the surface. (right) O--Si--O bond angles 
near the surface (dashed line) show broadening out to larger angles than in 
the bulk (solid line), indicating the presence of Si-3 species.
\label{fig:surf_defects}}
\end{figure}

\begin{figure}[h!]
\caption{Average density of water molecules perpendicular to the hydroxylated 
amorphous silica (left) and native oxide (right) surfaces. In both cases, 
the main peak is centred on the surface/water interface, with some water 
penetrating the surfaces and density oscillations continuing into the water layer. 
The water density in the centre of the simulation cells matches that of bulk 
water.
\label{fig:5}}
\end{figure}

\begin{figure}[h!]
\caption{(top) Force-separation curves for two hydroxylated amorphous 
silica (left) and native oxide (right) surfaces. The bare surfaces (dashed 
lines) repel (positive force) at separations below 0~\AA{}. Also plotted, 
from left to right, are curves corresponding to approximately 0.25, 0.50, 0.75, 1.00, 1.50, 
2.00 and 2.50~ML water coverage on each surface. The curves are translated 
upwards so that the force, at large separations, is zero. (bottom) Number of 
hydrogen bonds per water molecule as a function of surface separation for the 
amorphous silica (left) and native oxide (right) surfaces. Below about 1~ML (dotted 
lines), the number of hydrogen bonds does not reach the ideal surface density (upper 
dashed lines) before the surfaces repel. The maximum attractive force between  
surfaces at higher water coverages (solid lines) is limited by the hydrogen bond 
density between the surfaces, which tends to that of bulk water (lower dashed lines).
\label{fig:6}}
\end{figure}

\begin{figure}[h!]
\caption{Surface energy of two hydroxylated amorphous silica (solid line) and native 
oxide (dashed line) surfaces as a function of water coverage. The bonding energy of silica 
is approximately constant, with a small peak at 1~ML, and decreases at low coverages. 
The native oxide surfaces are more strongly bound at low coverages, before tending 
to the silica and bulk water surface energies at high coverages.
\label{fig:7}}
\end{figure}

\begin{figure}[h!]
\caption{Average water density profiles perpendicular to two natively oxidised 
Si wafers at equilibrium separation, at increasing water coverage. The 
corresponding force-separation curves are numbered (top left). At 0.5~ML 
coverage (1), a large force is required to separate the single density peak into two. 
At 1.0~ML (2), the bonding energy is a maximum, because a stable water network 
is set up between the two wafers. At 2.5~ML (3), the interfacial water 
density and associated surface energy is close to that of bulk water.}
\label{fig:8}
\end{figure}

\clearpage

\begin{center}
\includegraphics[width=3in]{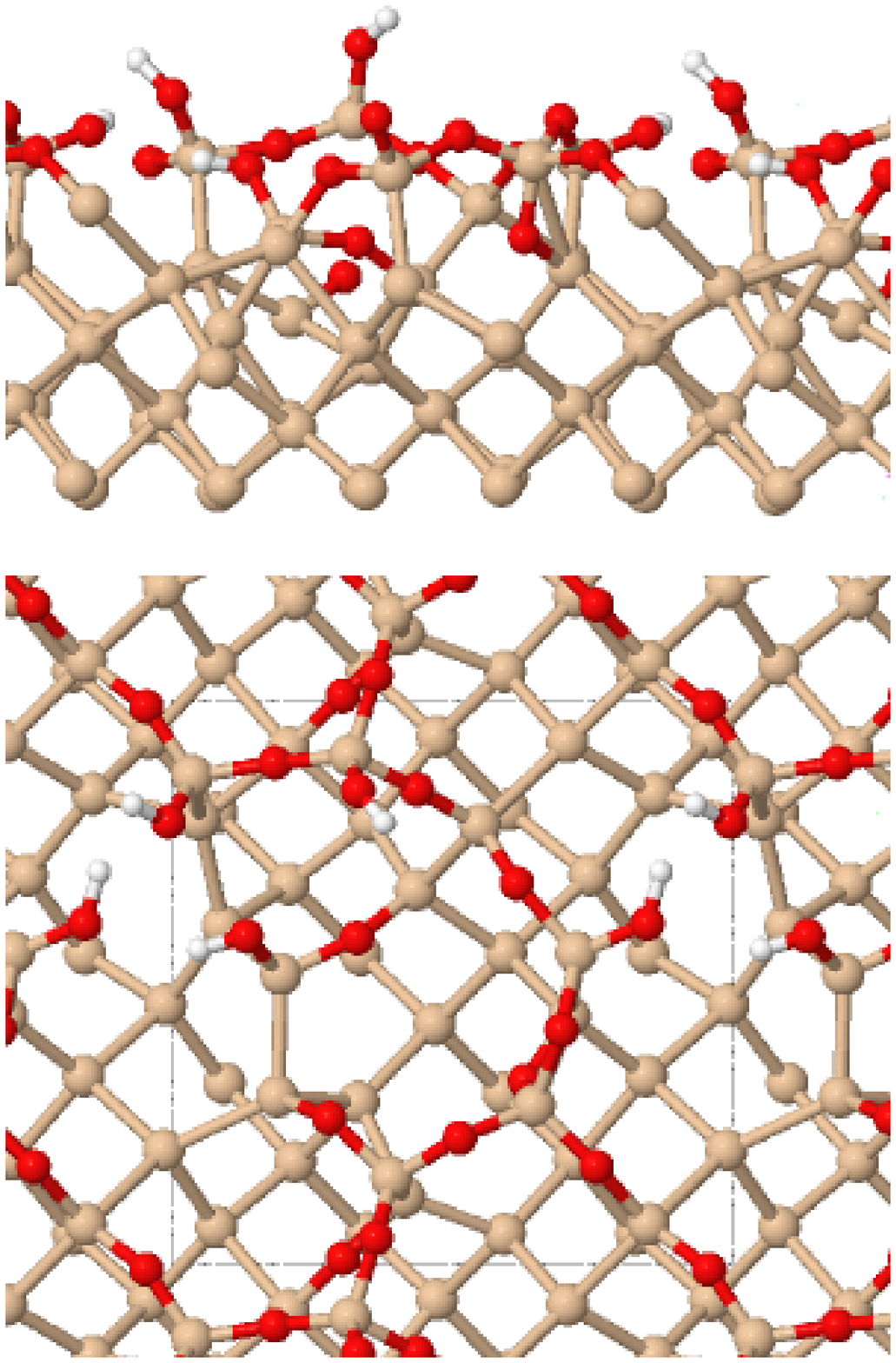}
\end{center}
\vfill\
D. J. Cole at al., Figure~\ref{fig:1}

\clearpage

\begin{center}
\includegraphics[width=3in]{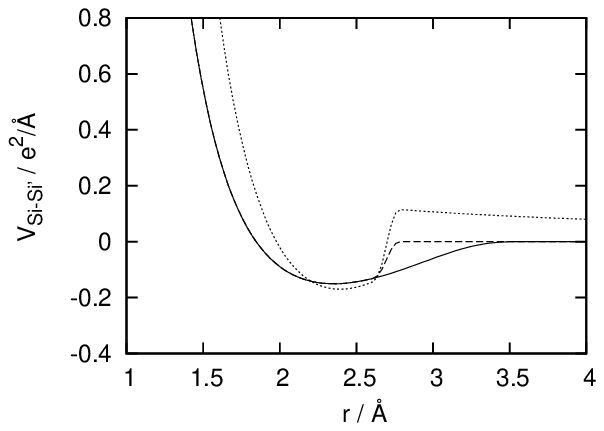}
\end{center}
\vfill\
D. J. Cole at al., Figure~\ref{fig:2}

\clearpage

\begin{center}
\includegraphics[width=3in]{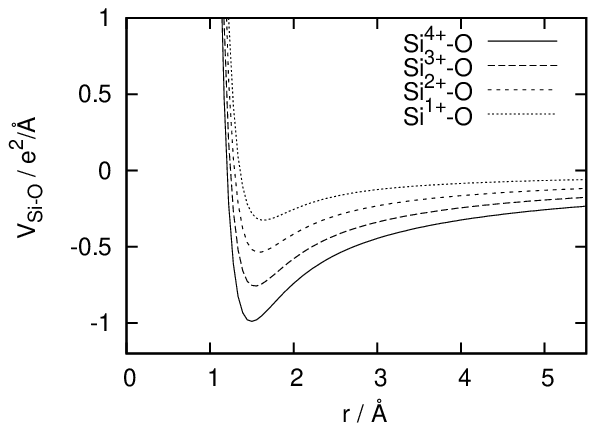}
\end{center}
\vfill\
D. J. Cole at al., Figure~\ref{fig:3}

\clearpage

\begin{center}
\includegraphics[width=5.9in]{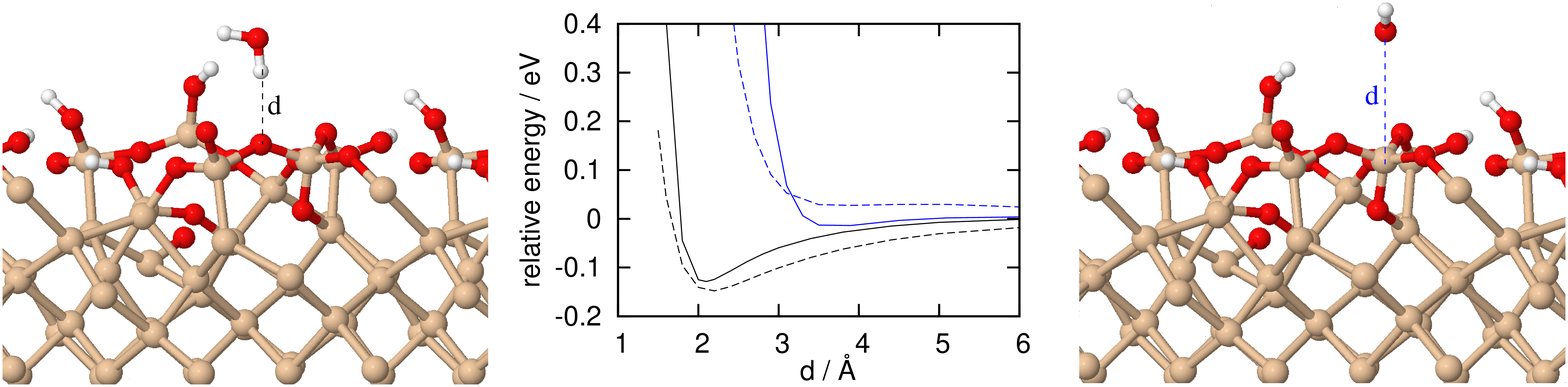}
\end{center}
\vfill\
D. J. Cole at al., Figure~\ref{fig:4}

\clearpage

\begin{center}
\includegraphics[width=5.9in]{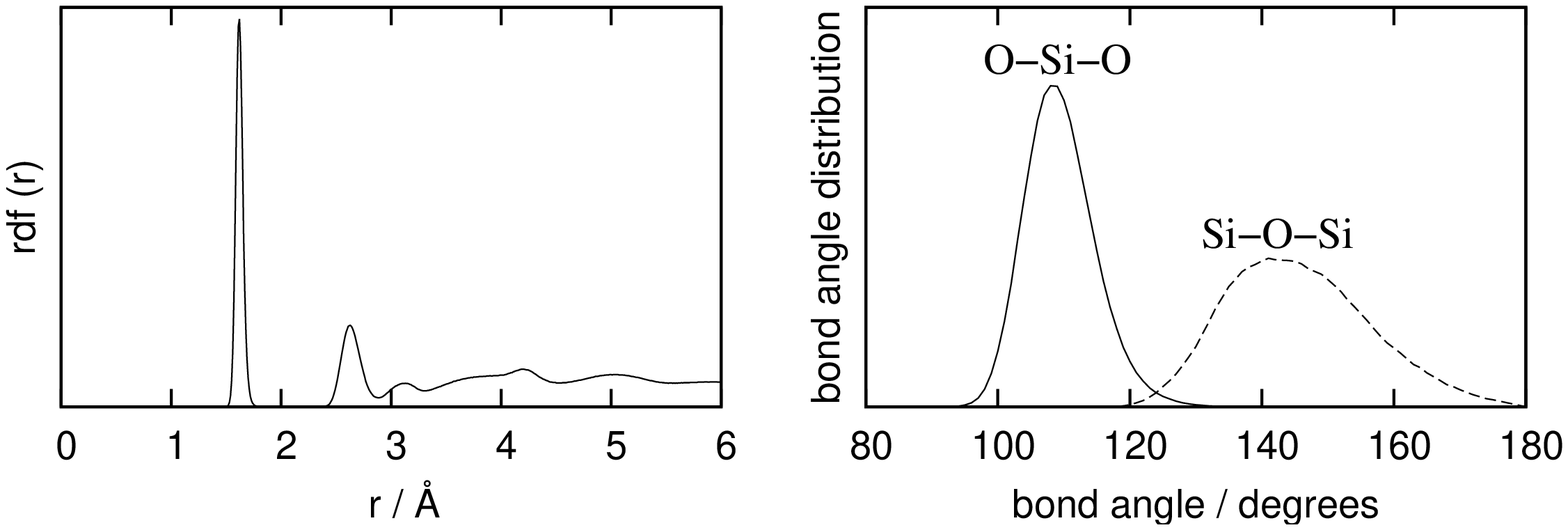}
\end{center}
\vfill\
D. J. Cole at al., Figure~\ref{fig:rdf}

\clearpage

\begin{center}
\includegraphics[width=5.9in]{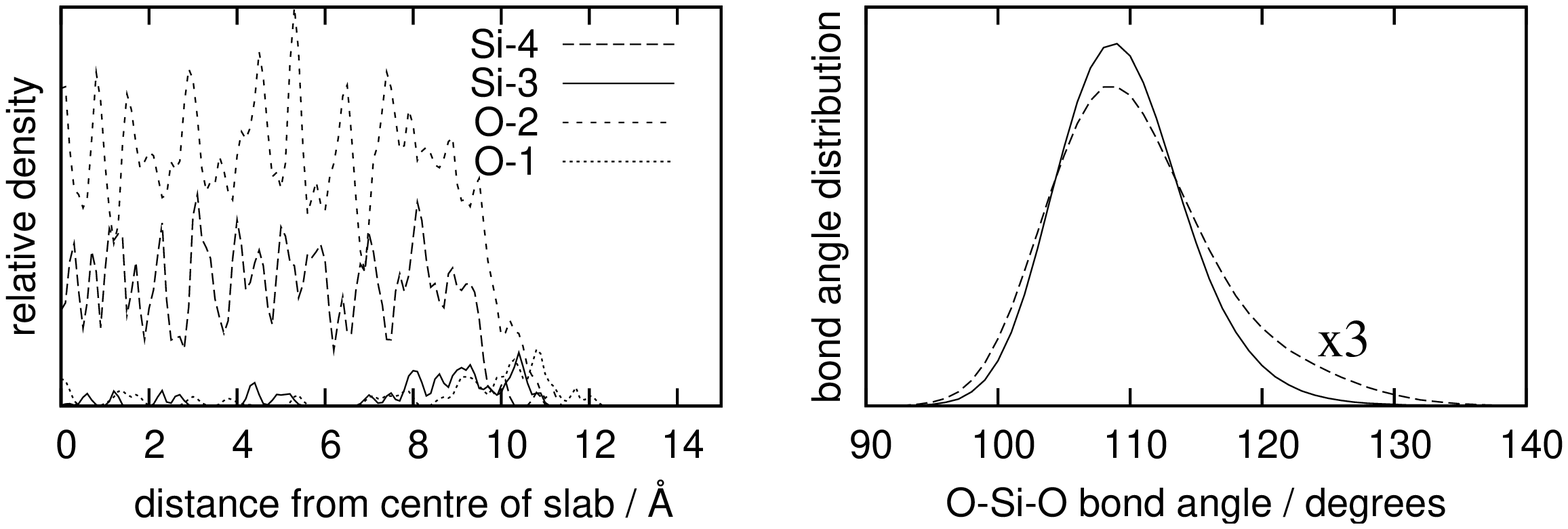}
\end{center}
\vfill\
D. J. Cole at al., Figure~\ref{fig:surf_defects}

\clearpage

\begin{center}
\includegraphics[width=5.9in]{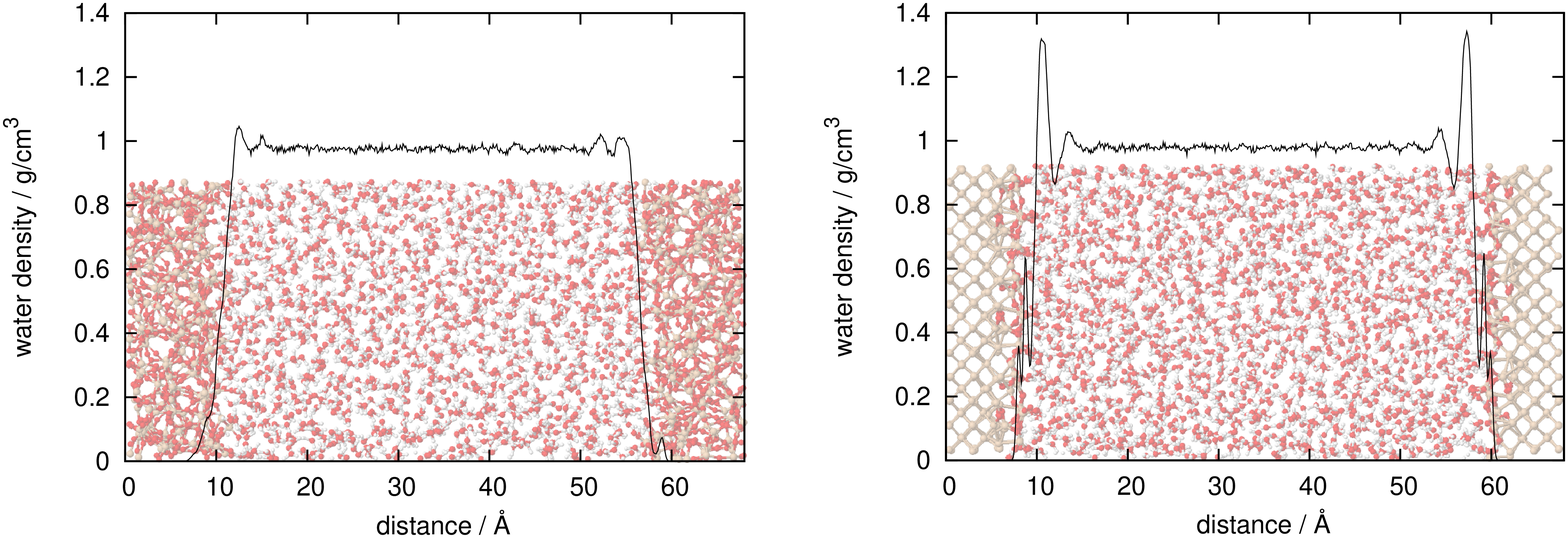}
\end{center}
\vfill\
D. J. Cole at al., Figure~\ref{fig:5}

\clearpage

\begin{center}
\includegraphics[width=5.9in]{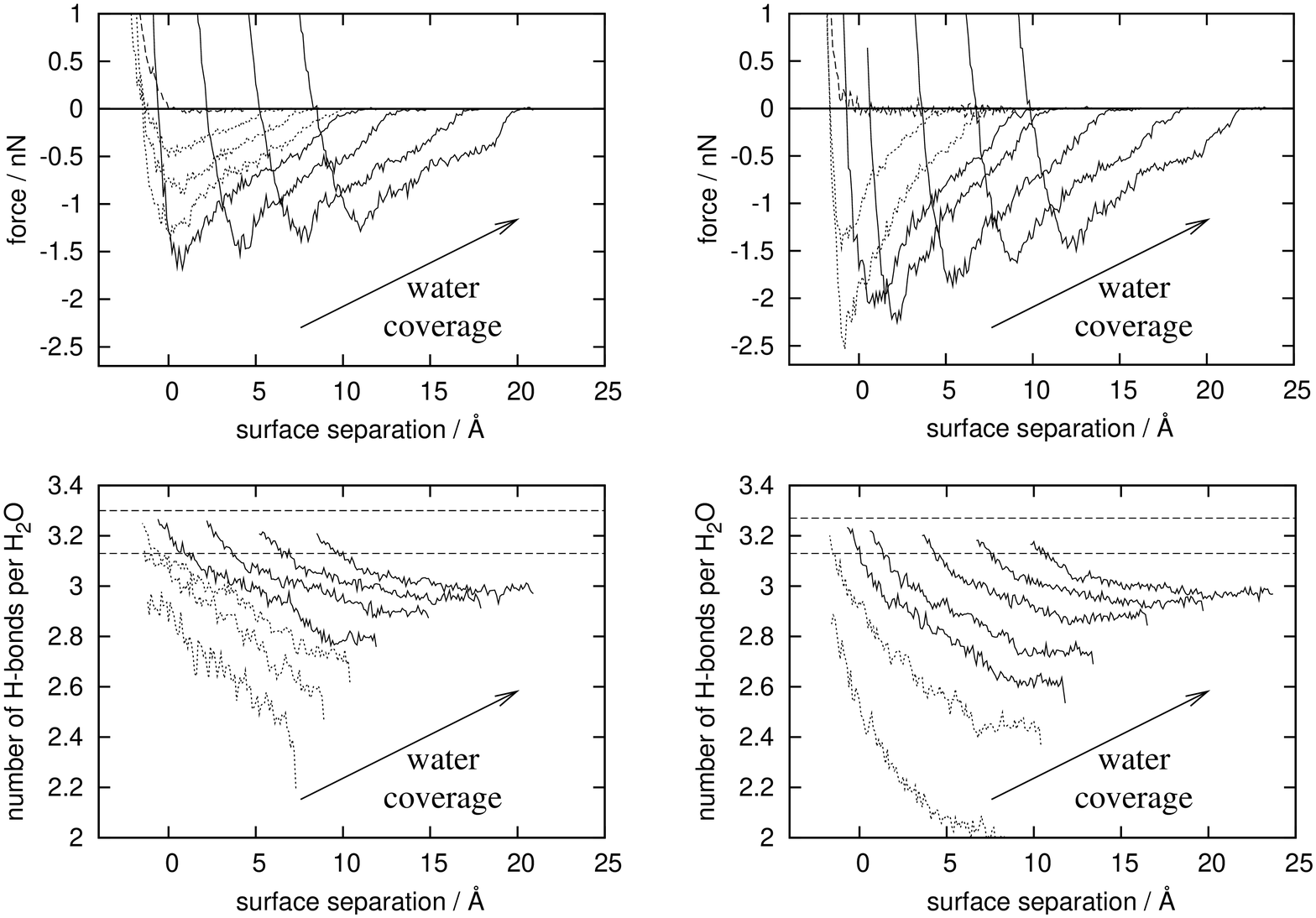}
\end{center}
\vfill\
D. J. Cole at al., Figure~\ref{fig:6}

\clearpage

\begin{center}
\includegraphics[width=4in]{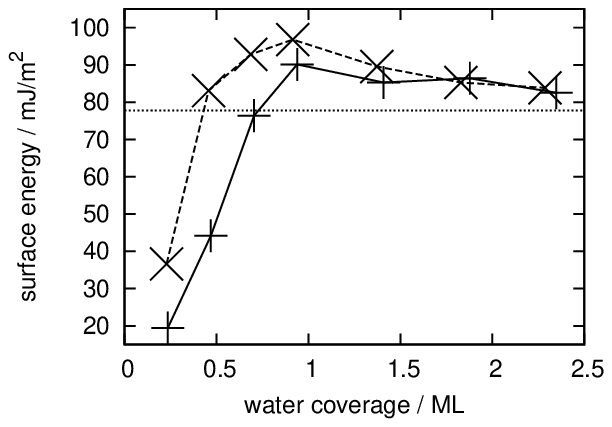}
\end{center}
\vfill\
D. J. Cole at al., Figure~\ref{fig:7}

\clearpage

\begin{center}
\includegraphics[width=5.5in]{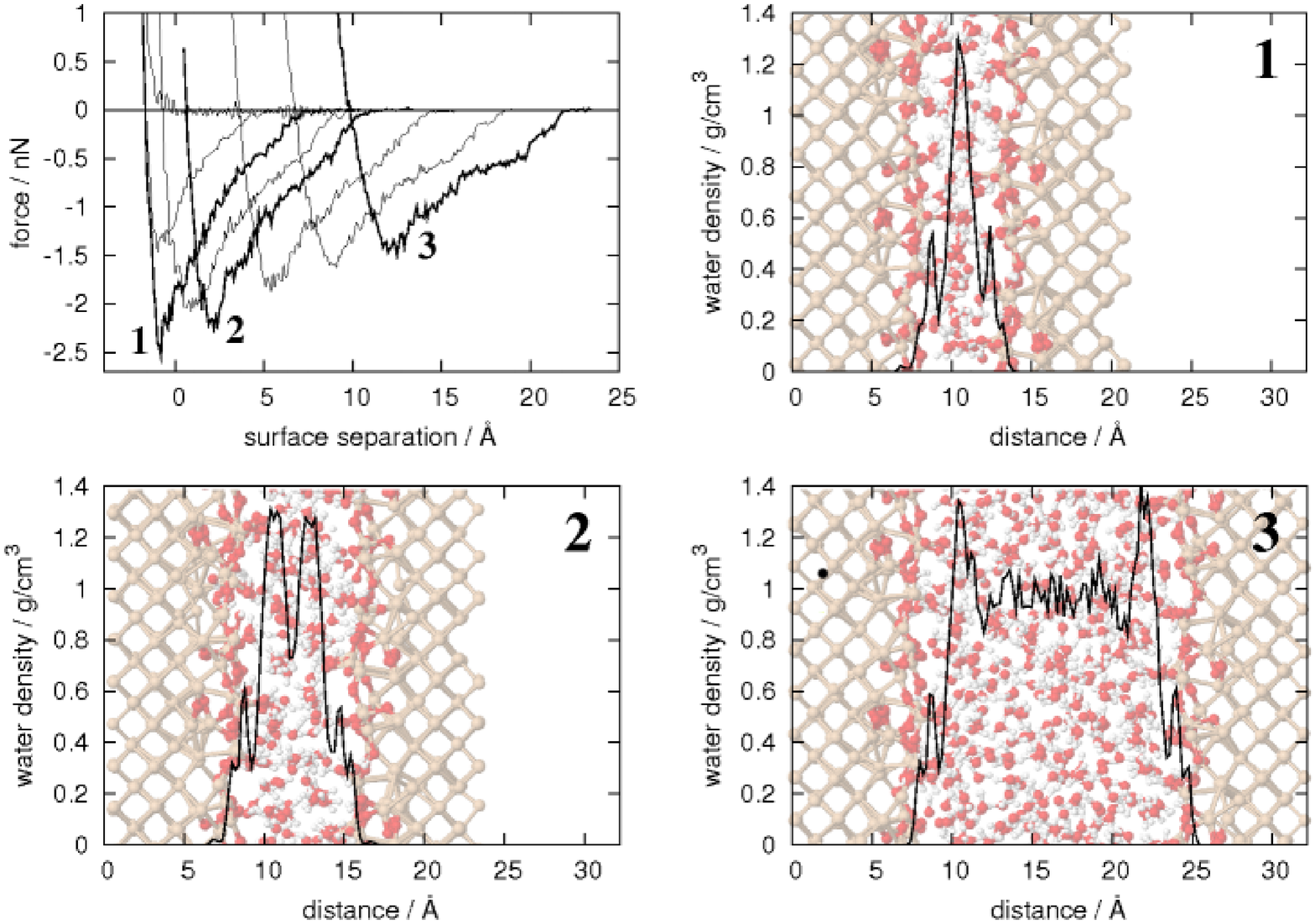}
\end{center}
\vfill\
D. J. Cole at al., Figure~\ref{fig:8}

\end{document}